\documentclass[12pt,a4paper]{conference}

\usepackage{fancyhdr}
\usepackage{graphicx,amsmath,amssymb,cite}
\usepackage{multind}
\makeindex{author} \makeindex{subject}

\newcommand{\beq}{\begin{equation}}
\newcommand{\eeq}[1]{\label{#1}\end{equation}}
\newcommand{\eeqn}{\end{equation}}

\newcommand{\beqa}{\begin{eqnarray}}
\newcommand{\eeqa}[1]{\label{#1}\end{eqnarray}}
\newcommand{\eeqan}{\end{eqnarray}}

\let\bar=\overbar

\newcommand{\Dslash}{\not{\hbox{\kern-4pt $D$}}}
\newcommand{\dslash}{\not{\hbox{\kern-2pt $\del$}}}

\newcommand{\msb}{{\bar{\ssstyle M \kern -1pt S}}}

\begin{document}

\Chapter{Scalar Mesons from an Effective Lagrangian Approach}
           {Scalar mesons}{A.H. Fariborz \it{et al.}}
\vspace{-6 cm}\includegraphics[width=6 cm]{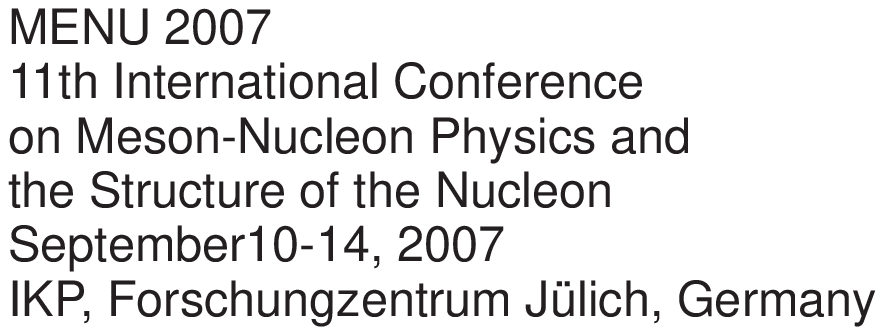}
\vspace{4 cm}

\addcontentsline{toc}{chapter}{{\it N. Author}} \label{authorStart}

\begin{raggedright}

{\it A.H. Fariborz
 \footnote{email:fariboa@sunyit.edu}
}\index{author}{Fariborz, A.H.}\\
Department of Mathematics/Physics\\
State University of New York Institute of Technology\\
P.O. Box 3050, Utica, New York 13504-3050, USA.
\bigskip

{\it R. Jora
 \footnote{email:cjora@phy.syr.edu}
}\index{author}{Jora, R.}\\
Physics Department\\
Syracuse University, Syracuse, New York 13244-1130, USA.
\bigskip

{\it J. Schechter
 \footnote{Speaker;email:schechte@phy.syr.edu}
}\index{author}{Schechter, J.}\\
Physics Department\\
Syracuse University, Syracuse, New York 13244-1130, USA.
\bigskip

\end{raggedright}

\begin{center}
\textbf{Abstract}
\end{center}
 A brief discussion of the recent interest in light
scalar mesons motivates the study of a generalized
linear sigma model. In an SU(3) flavor invariant
version of the model there is a prediction that the
the lighter scalars have sizeable ``four quark"
content. It is further predicted that one of the singlet
scalars should be exceptionally light.
Due to the presence of scalar mesons, 
the model gives ``controlled"
corrections to the current
algebra formula for threshold pion pion scattering.
These corections act in the direction to improve
agreement with current experiments.

\section{Introduction}
  A linear sigma model with both fields representing quark-antiquark 
type composites  and fields representing (in an unspecified configuration)
two quarks and two antiquarks, seems useful for
understanding the light
scalar spectrum of QCD.
To see this, note that, at present, the scalars below 1 GeV
appear to fit into a nonet as:
\begin{eqnarray}
I=0: m[f_0(600)]&\approx& 500\,\,{\rm MeV}                          
\nonumber \\
I=1/2:\hskip .7cm m[\kappa]&\approx& 800 \,\,{\rm MeV} 
\nonumber \\
I=0: m[f_0(980)]&\approx& 980 \,\,{\rm MeV} 
\nonumber \\
I=1: m[a_0(980)]&\approx& 980 \,\,{\rm MeV} 
\label{scalarnonet}
\end{eqnarray}
This level ordering may be compared with that of
the most conventional meson nonet- the low lying
vector mesons. In fact it is completely flipped compared to 
that one; the almost degenerate $I$=0 and $I$=1
vector states lie lowest rather than highest and the other
$I$=0 state lies highest. The ordering for vectors
 is conventionally understood
as due to the strange quark being heavier than the
 up and down quarks and the vectors being quark anti-quark 
composites. The ordering is gotten just by counting the number of
strange quarks in each state.
 It was pointed out a long time ago in Ref. \cite{j},
  using the same reasoning,
that the level order is automatically flipped
when the mesons are made of two quarks and two antiquarks
instead. That argument was given for a diquark- anti diquark
structure but is easily seen to also hold for the meson-
meson structure which was advocated for example
 in Ref. \cite{wi}. Thus, on empirical grounds a
four quark structure for the light scalars seems very 
suggestive and well worth investigating.

    Now one expects higher mass scalars related
  to p wave quark antiquark composites to also exist. It is natural
to expect mixing between states with the same quantum
numbers and there is some phenomenological evidence for
this as noted in Refs \cite{BFS3} and \cite{mixing}. Considering the 
importance of spontaneously broken chiral symmetry in low
energy QCD, it seems interesting to investigate packaging
 the whole scheme
in an effective chiral Lagrangian containing a ``two quark"
{\it chiral} nonet (with both scalars and pseudoscalars)
as well as a ``four quark" {\it chiral} nonet.

    We employ the 3$\times$3 matrix
chiral nonet fields;
\begin{equation}
M = S +i\phi, \hskip 2cm
M^\prime = S^\prime +i\phi^\prime.
\label{sandphi}
\end{equation}
Here $M$ represents scalar, $S$ and pseudoscalar,
$\phi$ quark-antiquark type states, while
$M^\prime$ represents states which are made of         
two quarks and two antiquarks.    The transformation
properties under  SU(3)$_{\rm L}\times$ SU(3)$_{\rm R}
\times$ U(1)$_{\rm A}$ are
\begin{equation}
M \rightarrow e^{2i\nu}
\, U_{\rm L} M U_{\rm R}^\dagger, \hskip 2cm
M^\prime \rightarrow e^{-4i\nu}
\, U_{\rm L} M^\prime U_{\rm R}^\dagger,
\label{Mchiral}
\end{equation}                          
where $U_{\rm L}$ and $U_{\rm R}$ are unitary unimodular
matrices, and the phase $\nu$ is associated with the
U(1)$_{\rm A}$ transformation, which also distinguishes
the two quark type from the four quark type fields. However,
together with our model, it
does not distinguish between different types of four quark
configuratons. That question is discussed in more detail
in ref. \cite{kknhh05}.
The general Lagrangian density which defines our model is
\begin{equation}
{\cal L} = - \frac{1}{2} {\rm Tr}
\left( \partial_\mu M \partial_\mu M^\dagger
\right) - \frac{1}{2} {\rm Tr}
\left( \partial_\mu M^\prime \partial_\mu M^{\prime \dagger} \right)
- V_0 \left( M, M^\prime \right) - V_{SB},            
\label{mixingLsMLag}
\end{equation}
where $V_0(M,M^\prime) $ stands for a function made
from SU(3)$_{\rm L} \times$ SU(3)$_{\rm R}$
(but not necessarily U(1)$_{\rm A}$) invariants
formed out of
$M$ and $M^\prime$. The quantity $V_{SB}$ stands for
chiral symmetry
breaking terms which transform in the same way
as the quark mass terms in
the fundamental QCD Lagrangian. The model was
proposed in section V of ref.\cite{BFMNS01} and 
followed up in refs. \cite{NR04}, \cite{FJS05}, 
\cite{1FJS07},\cite{2FJS07} and \cite{3FJS07},
the last two of which will be briefly described here.
Related models for thermodynamic properties of QCD
are discussed in refs. \cite{ythb}.
   In \cite{2FJS07}, we focused on general properties which
continued to hold
when $V_{SB}$ was set to zero while in \cite{3FJS07}
we included the SU(3) symmetric mass term:
\begin{equation}
V_{SB} = - 2\, A\, {\rm Tr} (S)
\label{vsb}
\end{equation}
where $A$ is a real parameter.

    A characteristic            
feature of the model is the presence of ``two-quark'' and
``four-quark'' condensates:
\begin{equation}
\left\langle S_a^b \right\rangle = \alpha_a \delta_a^b,
\quad \quad \left\langle S_a^{\prime b} \right\rangle =
\beta_a \delta_a^b.
\label{vevs}
\end{equation}
We shall assume the vacuum to be SU(3)$_{\rm V}$
invariant, which implies                                 
\begin{equation}
\alpha_1 = \alpha_2 = \alpha_3 \equiv \alpha, \hskip 2cm
\beta_1 = \beta_2 = \beta_3 \equiv \beta.
\end{equation}
The SU(3) particle content of the model consists of two
pseudoscalar octets, two pseudoscalar singlets, two scalar
octets  and two scalar singlets.   This gives us eight       
different masses and four mixing angles.                                                  

 Note that the transformation between the diagonal
fields ($\pi^+$ and $\pi'^+$)  and the
original pion fields is given as:
\begin{equation}
\left[
\begin{array}{c}  \pi^+ \\
                 \pi'^+
\end{array}
\right]
=
\left[
\begin{array}{c c}
                \cos\theta_\pi & -\sin \theta_\pi
\nonumber               \\
\sin \theta_\pi & \cos \theta_\pi
\end{array}
\right]
\left[
\begin{array}{c}
                        \phi_1^2 \\
                        {\phi'}_1^2
\end{array}
\right].
\label{mixingangle}
\end{equation}                                     
Thus 100 $\sin^2\theta_\pi$ represents the four quark
percentage of the ordinary pion while 100 $\cos^2\theta_\pi$
represents the four quark percentage of the ``heavy"
pion in the model.
Also of relevance is the one particle
piece of the isovector axial vector current,
\begin{equation}
(J_\mu^{axial})_1^2=F_\pi\partial_\mu\pi^+
+F_{\pi'}\partial_\mu\pi'^+ +\cdots,
\label{axcur}
\end{equation}
where,
\begin{eqnarray}
F_\pi&=&2\, \alpha \cos\theta_\pi -2\, \beta \sin\theta_\pi,
\nonumber \\
F_{\pi'}&=&2\, \alpha \sin\theta_\pi + 2\, \beta \cos \theta_\pi.
\label{dconstants}
\end{eqnarray}                                                       
Note that $\tan\theta_\pi=-\beta/\alpha$ when the pion is massless.

\section{Specific Lagrangian}
As discussed in ref.\cite{2FJS07} one
 may obtain certain general results
from tree level Ward identities without
 restrictions on the form of
the potential $V_0$. For a complete description however,
a specific form must be furnished.
The leading choice of terms corresponding
to eight or fewer quark plus antiquark lines at
 each effective underlying vertex
reads\cite{2FJS07}:
\begin{eqnarray}
V_0 =&-&c_2 \, {\rm Tr} (MM^{\dagger}) +
c_4^a \, {\rm Tr} (MM^{\dagger}MM^{\dagger})
\nonumber \\
&+& d_2 \,
{\rm Tr} (M^{\prime}M^{\prime\dagger})
     + e_3^a(\epsilon_{abc}\epsilon^{def}M^a_dM^b_eM'^c_f + h.c.)
\nonumber \\
     &+&  c_3\left[ \gamma_1 {\rm ln} (\frac{{\rm det} M}{{\rm det}
M^{\dagger}})
+(1-\gamma_1)\frac{{\rm Tr}(MM'^\dagger)}{{\rm Tr}(M'M^\dagger)}\right]^2.
\label{SpecLag}
\end{eqnarray}
     All the terms except the last two have been chosen to also
possess the  U(1)$_{\rm A}$
invariance. Further details of the U(1)$_{\rm A}$ aspect are given in
ref.\cite{2FJS07}.                      

As the corresponding experimental inputs
\cite{ropp} we take the non-strange
quantities:
\begin{eqnarray}
m(0^+ {\rm octet}) &=& m[a_0(980)] = 984.7 \pm 1.2\, {\rm MeV}
\nonumber \\
m(0^+ {\rm octet}') &=& m[a_0(1450)] = 1474 \pm 19\, {\rm MeV}
\nonumber \\
m(0^- {\rm octet}') &=& m[\pi(1300)] = 1300 \pm 100\, {\rm MeV}
\nonumber \\
m(0^- {\rm octet}) &=& m_\pi = 137 \, {\rm MeV}
\nonumber \\
F_\pi &=& 131 \, {\rm MeV}
\label{inputs1}
\end{eqnarray}                                                    
Clearly m[$\pi$(1300)] has a large uncertainty and will essentially
be regarded as a free parameter.

\begin{figure}[h]
\begin{center}
\vskip 1cm
\includegraphics[width=8 cm]{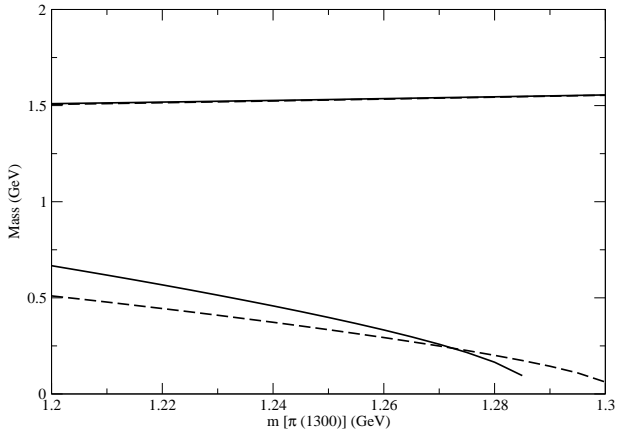}
\hspace{4 cm}
\caption{
The predictions for the masses of the
two SU(3) singlet scalars
vs. $m[\pi(1300)]$. The solid lines correspond to the massive
pion case while the dashed lines correspond to
the massless pion case.                  
} \label{ms0vsmpip}
\end{center}
\end{figure}

\begin{figure}[h]
\begin{center}
\vskip 1cm
\includegraphics[width=8 cm]{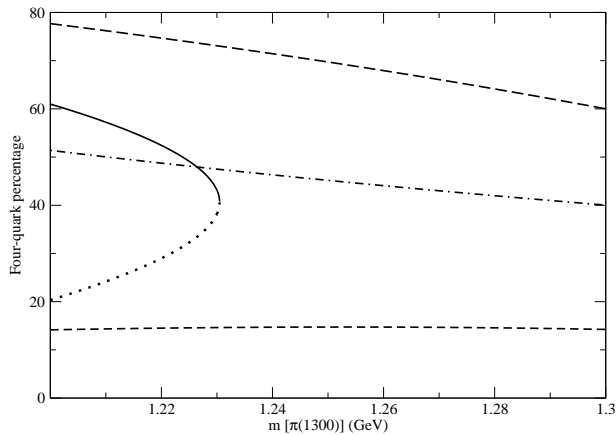}
\hspace{4 cm}
\caption{
Four quark percentages of the pion (dashed line), the $a_0(980)$
 (top long-dashed line), the very light $0^+ {\rm singlet}$ (dotted-dashed
line)  and the
$\eta(958)$ in the scenario where the higher state is identified as the
$\eta(1475)$ (curve containing both solid and dotted pieces)
as functions of the undetermined input parameter,
$m[\pi(1300)]$. Note that there are two solutions for the $\eta(958)$: the
dotted curve choice
gives it a predominant two quark structure and the solid curve choice,
a larger four quark content.                                           
} \label{fqcvsmpip}
\end{center}
\end{figure}

    Predictions for the two unspecified 0$^+$ masses
are shown in Fig.\ref{ms0vsmpip}. The lighter of these clearly invites
us to identify it as the sigma.
    Predictions for the four quark contents of the 
lightest four SU(3) multiplets are shown in Fig.\ref{fqcvsmpip}.
Clearly the lighter 0$^-$ octet is primarily two quark and the
lighter  0$^-$
singlet has a primarily two quark solution. On the other hand,
both the lighter 0$^+$ octet and singlet have large
four quark content!

\section{Pion pion scattering}
    The tree level form for the conventional Mandelstam
amplitude in the present model is:
\begin{equation}
A(s,t,u)=-\frac{g}{2}+
\sum_i\frac{g_i^2}{m_i^2-s},
\label{Afirst}
\end{equation}
where $g$ is the four pion coupling constant, the 
four $g_i$'s are the three-point coupling constants
of two pions with each of the four scalar isosinglets
and the $m_i$ represent the masses of the four scalar
isosinglets. To really understand what is happening
 we should expand
in powers of $(s-m_{\pi}^2)$:
\begin{eqnarray}
A(s,t,u)&=&-\frac{g}{2}+
\sum_i\frac{g_i^2}{m_i^2-m_\pi^2}\left[
1+\frac{s-m_\pi^2}{m_i^2-m_\pi^2}+
(\frac{s-m_\pi^2}{m_i^2-m_\pi^2})^2+\cdots\right]
\nonumber \\
&\approx&(s-m_\pi^2)\left[\frac{2}{F_\pi^2}
+(s-m_\pi^2)\sum_i\frac{g_i^2}{(m_i^2-m_\pi^2)^3}+\cdots\right].
\label{correction}
\end{eqnarray}
The exact first equation contains, for each $m_i$,
a geometrical expansion in the quantity $(s-m_\pi^2)/
(m_i^2-m_\pi^2)$. Thus the radius of convergence
in s for this expression is the squared mass of the lightest
scalar isosinglet.
 To apply this expression in the
resonance region we must, of course, unitarize the
formula in some way. Here we will look                
at the threshold region.
In going from the first to the second equation
of Eq.(\ref{correction}) we used
the facts established 
in ref.\cite{2FJS07} for the $m_{\pi}=0$ case
and in ref.\cite{3FJS07}
(in the very good approximation where $F_{\pi'}=0$)
for the massive pion case that:
1) the sum of the first two
terms of the first equation vanishes
and 2) the third term
of the first equation simplifies
to becomes the first,
current algebra, term                              
of the second equation.
These results hold for any chiral symmetric choice of the 
potential. The third term
of the second
equation represents the
model dependent leading correction
to the usual current algebra
formula. It depends on the masses of the
scalar mesons and would
vanish in a hypothetical limit (often used)
in which the scalar
meson masses are taken to infinity. At low
energies the third term is seen to be  suppressed
by order $(m_{\pi}/m_i)^2$ compared to the current
algebra term.

 For comparison, we give the usual current
algebra results \cite{W} for the two s-wave
scattering lengths:
\begin{eqnarray}
m_\pi a_0^0 &=& \frac{7m_\pi^2}{16{\pi}F_\pi^2}
\approx 0.15,
\nonumber \\
m_\pi a_0^2 &=& \frac{-2m_\pi^2}{16{\pi}F_\pi^2}
\approx -0.04.
\label{cascatlen}
\end{eqnarray}                                       
 Recent experimental data on the s-wave scattering
lengths $a_0^0$ and $a_0^2$ include the following,

NA48/2 collaboration \cite{NA48}:
\begin{equation}
m_{\pi^+}(a_0^0-a_0^2)=0.264 \pm 0.015
\end{equation}
\begin{equation}
m_{\pi^+}a_0^0 = 0.256 \pm 0.011
\end{equation}

E865 Collaboration \cite{E865}:
\begin{equation}
m_{\pi^+}a_0^0 = 0.216 \pm 0.015
\end{equation}

DIRAC Collaboration \cite{dirac}
\begin{equation}
m_{\pi^+}a_0^0 =0.264_{-0.020}^{+0.038}               
\end{equation}

\begin{figure}
\begin{center}
\vskip 1cm
\includegraphics[width=8 cm]{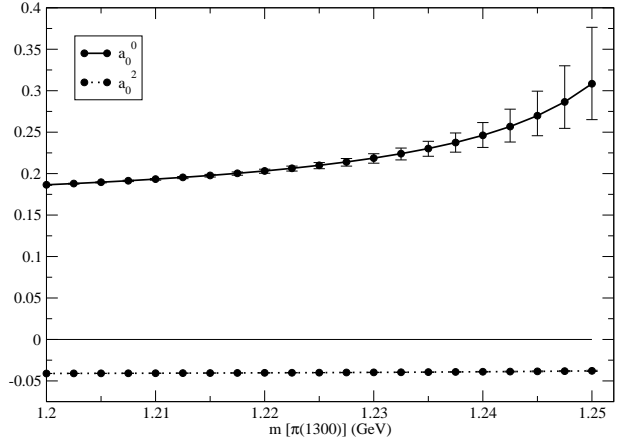}
\hspace{4 cm}
\caption{Top curve: $I=J=0$ scattering length, $m_{\pi}a_0^0$ vs. 
$m[\pi(1300)]$.
Bottom curve: $I=2$, $J=0$ scattering length, $m_{\pi}a_0^2$ vs.
$m[\pi(1300)]$.
   The error bars reflect the
uncertainty of $m[a_0(1450)]$.         
} \label{scatle}
\end{center}
\end{figure}

Clearly, the current algebra value of $a_0^0$ is lower
than experiment. Including the corrections from non-infinite
scalar meson masses with the choice of $V_0$ in Eq.(\ref{SpecLag}),
yields the results shown in Fig.\ref{scatle}.
 The corresponding lightest scalar mass for each value of
m[$\pi$(1300)] can be read from Fig. \ref{ms0vsmpip}. The
value of $a_0^2$ for the non resonant channel is not altered
much but it can be seen that the prediction for $a_0^0$ is
definitely improved.

\section{Further discussion}
\begin{figure}
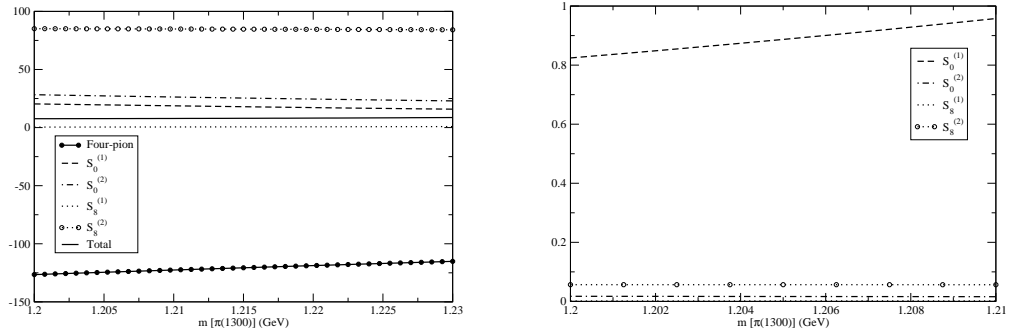

\begin{center}
\vskip 1cm
\includegraphics[width=6 cm]{fig7.eps}
\hskip 1cm
\includegraphics[width=6 cm]{fig8.eps}
\caption{Left: Individual contributions in 
Eq.(\ref{Afirst}) to 
$A(s,t,u)$
at threshold. Right: Individual contributions to the second of
 Eq.(\ref{correction})
to $A(s,t,u)$ at threshold.
} \label{controlled}
\end{center}
\end{figure}

1. A criticism of linear sigma models is that
they obtain the current algebra result
 as an almost complete
cancellation of large quantities.
 This may be seen in the 
left panel of Fig. \ref{controlled}
 which shows the five different
terms in Eq.(\ref{Afirst}) appearing to almost cancel
 in a haphazard pattern. On the other hand, once
we make the Taylor expansion in Eq.(\ref{correction}), the
corrections to the current algebra result are seen
in the right panel to be 
completely dominated by the lightest scalar; this
contribution is suppressed, as noted above, by
$(m_\pi/m_\sigma)^2$.

2. We plan to increase the accuracy of our model by including
SU(3) breaking effects and also by unitarizing the model
so that it would be suitable also in the
 energy region around the scalar 
resonances. The simplest unitarization is the K-matrix type.
It was caried out for the two flavor linear sigma
 model in ref.\cite{AS}.
We expect the results for the present model to
 be generally similar
to the ones obtained in ref.\cite{BFMNS01}
 for the single $M$ three flavor linear model.

3. A possible general question about the present
model is that it introduces both states made of a quark
and an antiquark as well as states with two quarks
and two antiquarks. According to the usual 't Hooft large $N_c$
extrapolation \cite{th} of QCD the ``four quark" states
are expected to be suppressed. However, it was recently
pointed out \cite{ss07} that the alternative, mathematically
allowed, Corrigan Ramond \cite{cr} extrapolation does not
 suppress the
multiquark states. This kind of extrapolation may be   
relevant for understanding the physics of the light scalar mesons. 
\section*{Acknowledgments}

We are happy to thank A. Abdel-Rehim, D. Black, M. Harada,
S. Moussa, S. Nasri and F. Sannino for many helpful
related discussions.
The work of A.H.F. has been partially supported by the
NSF Award 0652853.
The work of R.J. and J.S. is supported in part by the U. S. DOE under
Contract no. DE-FG-02-85ER 40231.


\end{document}